\documentclass{webofc}
\usepackage[varg]{txfonts}   
\newcommand{\be}{\begin{equation}}
\newcommand{\ee}{\end{equation}}

\newcommand\ba{\begin{eqnarray}}
\newcommand\ea{\end{eqnarray}}
\begin{document}
\title{
Renormdynamics
and Generalized Analytic Functions
}
%
%

\author{
Nugzar
\lastname{Makhaldiani}\inst{1
}\fnsep\thanks{\email{
mnv@jinr.ru}} 
}

\institute{
Joint Institute for Nuclear Research\\
 Joliot-Curie 6,
Dubna, Moscow region,
Russia, 141980
          }

\abstract{
We provide definitions of renormdynamic motion equations and some properties of renormdynamic functions with examples. Formal, longwave and shortwave solutions of the canonical equation for generalized (pseudo) analytic functions (GPF) are considered.

}
\maketitle

We say that we find {\bf New Physics} (NP) when either we find a phenomenon which is forbidden by Standard Model (SM) in principal - this is the qualitative level of NP - or we find a significant deviation between precision calculations in SM of an observable quantity and a corresponding experimental value.

In Quantum field theory (QFT) existence of a given theory means, that we can control its
behavior at some scales (short or large distances) by
renormalization theory \cite{Collins}. If the theory exists, than
we want to solve it, which means to determine what happens on
other
scales. This is the problem (and content)
of {\bf Renormdynamics} \cite{Makhaldiani17}. The result of the Renormdynamics, the solution
of its discrete or continual motion equations, is the effective QFT
on a given scale (different from the initial one).

We will call Renormdynamic Functions (RDF)  functions $g_n=f_n(t)$ which are solutions of the RD motion equations
\ba
\dot{g}_n=\beta_n(g), 1\leq n\leq N.
\ea
In the simplest case of one coupling constant (e.g. in Quantum electrodynamics (QED), Quantum chromodynamics (QCD)) the function $g=f(t)$
is constant, $g=g_c$ when $\beta(g_c)=0,$ or
is invertible (monotone). Indeed,
\ba
\dot{g}=f'(t)=f'(f^{-1}(g))=\beta(g).
\ea
Each monotone interval ends by Ultraviolet (UV) and Infrared (IR) fixed points and describes
corresponding phase of the system. Note that the simplest case of the classical dynamics, the Hamiltonian system with one degree of freedom, is already two dimensional, so we have
no analog of one charge renormdynamics.

There are different  parameterizations $a=f(A).$ The values of the critical points in different parameterizations differ, but the scale is same:
\ba
\dot{a}=b(a)=f'B(A),\ b(a)=0\Leftrightarrow B(a)=0,\ f'\neq0.
\ea
 In general case of $N$ coupling constants we have the similar result,
\ba
&&a_n=f_n(A_1,A_2,...A_N),\ 1\leq n\leq N,\cr
&&\dot{a}_n=b_n(a)=f'_{nm}B_m(A),\ b_n(a)=0\Leftrightarrow B_n(a)=0,\ detf'\neq0,\ f'_{nm}=\frac{\partial f_n}{\partial A_m}.
\ea
This is important observation because it helps us not only identify such an important quantities as phase transition temperature, hadronization scale, valence quark scale,..., but also control quality of parametrization and systematic errors of approximations.

In string theory, the connection between conformal invariance of the effective theory on the parametric world sheet and the motion equations of the fields on the embedding space is well known \cite{GreenString}, \cite{Ketov}.
A more recent topic in this direction is AdS/CFT Duality \cite{Maldacena}.
In this approach for QCD coupling constant the following expression was obtained  \cite{Brodsky}
\ba
\alpha_{AdS}(Q^2)=\alpha(0)e^{-Q^2/4k^2}.
\ea
A corresponding  $\beta$-function is
\ba
\beta(\alpha_{AdS})=\frac{d\alpha_{AdS}}{\ln Q^2}=-\frac{Q^2}{4k^2}\alpha_{AdS}(Q^2)
=\alpha_{AdS}(Q^2)\ln\frac{\alpha_{AdS}(Q^2)}{\alpha(0)}
\ea
So, this renormdynamics of QCD interpolates between the IR fixed point $\alpha(0),$ which we take as  $\alpha(0)=2,$ and the UV fixed point
$\alpha(\infty)=0.$

For the QCD running coupling considered in \cite{Diakonov}
\ba
\alpha(q^2)=\frac{4\pi}{9\ln(\frac{q^2+m_g^2}{\Lambda^2})},
\ea
where $m_g=0.88 GeV,\ \Lambda=0.28 GeV,$
the $\beta-$function of renormdynamics is
\ba\label{mgb}
&&
\beta(\alpha)=-\frac{\alpha^2}{k}(1-c\exp(-\frac{k}{\alpha})),
\cr
&&
k=\frac{4\pi}{9}=1.40,\ c=\frac{m_g^2}{\Lambda^2}=(3.143)^2=9.88,
\ea
for a nontrivial (IR) fixed point we have
\ba
\alpha_{IR}=k/\ln c
=0.61
\ea
For $\alpha(m)=2,$ at valence quark scale $m$ we predict the gluon (or valence quark) mass as
\ba
m_g=\Lambda e^{\frac{k}{2\alpha(m)}}=1.42\Lambda=
m_N/3,\
\Lambda=220 MeV.
\ea
From the nonperturbative $\beta-$functions we see that besides perturbative phase, with asymptotic freedom, there is also nonperturbative phase with infrared fixed point and rising coupling constant at higher energies. At small scales in QCD we have perturbative, small coupling, phase and nonperturbative, strong coupling, phase. The phases unify at the IR fixed point beyond of which we have hadronic phase.

It is nice to have a nonperturbative $\beta-$function like (\ref{mgb}), but it is more important to see which kind of nonperturbative corrections we need to have a phenomenological coupling constant dynamics.
It was noted \cite{Voloshin} that in valence quark parametrization $\alpha_s(m)=2,\ $at a valence quark scale $m.$

The theory of analytic functions of a complex variable occupies a central place in analysis.
  Riemann considered the unique continuation property to be the most characteristic feature of analytic functions.
GPF do possess the unique continuation property, and each class of
GPF has almost as much structure as the class of analytic functions. In particular, the operations of complex differentiation and complex integration have meaningful counterparts in the theory of GPF
and this theory generalizes not only the Cauchy-Riemann approach to function theory but also that of Weierstrass.
Such functions were considered by Picard  and by Beltrami, but the first significant result was obtained by Carleman in 1933, and a systematic theory was formulated
by Lipman Bers \cite{Bers}
and Ilia Vekua (1907-1977) \cite{Vekua}.
For more resent results see \cite{Giorgadze}.

Analytic function $f=u+iv$ satisfy the partial differential equation
 $\partial_{\overline{z}}f=0,$
where complex differential operators are defined as
\ba
\partial_{\overline{z}}=\frac{\partial}{\partial \overline{z}}:=\frac{1}{2}(\partial_x+i\partial_y),\ \partial_{z}=\frac{\partial}{\partial z}:=\frac{1}{2}(\partial_x-i\partial_y)
\ea
Generalized analytic functions $f=u+iv$  satisfy the following
generalized Cauchy-Riemann equation \cite{Vekua}
\ba
\partial_{\overline{z}}f=Af+B\bar{f}+J,\ A=A_0+iA_1,\ B=B_0+iB_1,\ J=j_1+ij_2
\ea
or in terms of the real $u$ and imaginary $v$ components canonical form of the elliptic systems of partial differential equations of the first order
\ba
&&u_x-v_y=au+bv+j_1,\ a=A_0+B_0,\ b=-A_1+B_1,\cr
&&u_y+v_x=cu+dv+j_2,\ c=A_1+B_1,\ d=A_0-B_0,
\ea
or in matrix form
\ba\label{EGPF}
&& D\psi=E\psi+J,\ D=\left(
              \begin{array}{cc}
                \partial_x &-\partial_y \\
                \partial_y & \partial_x \\
              \end{array}
            \right)=\partial_x-i\sigma_2\partial_y,\cr
            && E=\left(
                                                       \begin{array}{cc}
                                                         a & b \\
                                                         c & d \\
                                                       \end{array}
                                                     \right),\
            \psi=\left(
                                                            \begin{array}{c}
                                                              u \\
                                                              v \\
                                                            \end{array}
                                                          \right),\ J=\left(
                                                            \begin{array}{c}
                                                              j_1 \\
                                                              j_2 \\
                                                            \end{array}
                                                          \right).
\ea

In the classical sense by a solution of the system of equations (\ref{EGPF}) we understand a pair of real continuously differentiable functions $u(x,y),\ v(x,y)$ of the real variables $x$ and $y$ which satisfy this system everywhere in a domain $G$.  Such solutions, however, exist only for a comparatively narrow class of equations.

The formal solution of the canonical equation for GPF (\ref{EGPF}) is
\ba
\psi=\psi_0+RJ,\ R=(D-E)^{-1},\ (D-E)\psi_0=0.
\ea
Let us introduce a
length parameter $l=h^{-1}$, which is of order of the source $J$ size,\ $x_n\rightarrow lx_n.$
Then, for the resolvent $R,$
we will have the longwave and shortwave expansions,
\ba
&&R_{LW}:=(lD-E)^{-1}=-E^{-1}\sum_{n\geq0}l^n(DE^{-1})^n,\cr
&&R_{ShW}:=(lD-E)^{-1}=hD^{-1}\sum_{n\geq0}h^n(ED^{-1})^n,\cr
&&E^{-1}=\left(
           \begin{array}{cc}
             d & -b\\
             -c & a \\
           \end{array}
         \right)/\Delta_E,\ \Delta_E=ad-bc,\cr
         &&D^{-1}=\Delta_D^{-1}\left(\begin{array}{cc}
                    \partial_x & \partial_y \\
                    -\partial_y & \partial_x
                  \end{array}
                  \right)=\Delta_D^{-1}(\partial_x+i\sigma_2\partial_y),\ \Delta_D=\partial_x^2+\partial_y^2
\ea
There is a fairly complete theory of generalized analytic functions; it represents an essential extension of the classical theory preserving at the same time its principal features \cite{Vekua}.

\end{document}